\documentclass[conference]{IEEEtran}

\usepackage[english]{babel}
\usepackage{amsmath}
\usepackage{amsfonts}
\usepackage[amsmath]{ntheorem}
\usepackage{graphicx}
\usepackage[nolist]{acronym}
\usepackage{blkarray}
\usepackage{color}
\usepackage{amssymb}
\usepackage{graphicx}
\usepackage{tikz}
\usetikzlibrary{positioning}
\usetikzlibrary{calc,bending}
\usepackage{babel,blindtext}
\usepackage[capitalize]{cleveref}
\usepackage{algorithm}
\usepackage{algpseudocode}
\usepackage{bbm}
\usepackage[normalem]{ulem}
\usepackage{mathtools}
\usepackage{balance}

\acrodef{ai}[AI]{Artificial Intelligence}
\acrodef{5g}[5G]{Fifth Generation}
\acrodef{ml}[ML]{Machine Learning}
\acrodef{ib}[IB]{Information Bottleneck}
\acrodef{dib}[DIB]{Deterministic Information Bottleneck}
\acrodef{rl}[RL]{Reinforcement Learning}
\acrodef{mdp}[MDP]{Markov Decision Process}
\acrodef{rd}[RD]{Rate Distortion}
\acrodef{marl}[MARL]{Multi-Agent Reinforcement Learning}
\acrodef{nn}[NN]{Neural Network}
\acrodef{ot}[OT]{Optimal Transport}
\acrodef{snr}[SNR]{Signal to Noise Ratio}
\acrodef{psnr}[PSNR]{Peak Signal to Noise Ratio}

\newcommand{\enc}{\lambda}                                          
\newcommand{\dec}{\gamma}                                           
                                          
\newcommand{\mt}{T}                                                 
\newcommand{\mct}{\mathcal{T}}                                      
\newcommand{\cb}{\mathcal{C}}                                       
\newcommand{\q}{Q}                                                  
                                               
\newcommand{\obs}{o}                                                
                                               
\newcommand{\sem}{x}                                                
                                      
\newcommand{\act}{a}                                                
                                               
\newcommand{\sspace}{\mathcal{X}}                                   
\newcommand{\ospace}{\mathcal{O}}                                   
\newcommand{\aspace}{\mathcal{A}}                                   
\newcommand{\lin}{l}                                                            
\newcommand{\ls}{s}                                                             
\newcommand{\lt}{t}                                                            
\newcommand{\lang}{\ell}                                               
\newcommand{\satom}{P}                                              
\newcommand{\satomsource}[1]{P_{#1}^{\ls}}                     
\newcommand{\satomtarget}[1]{P_{#1}^{\lt}}                      
\newcommand{\infotransfer}[2]{\rho_{\satomsource{#1}\xrightarrow{}\satomtarget{#2}}}

\newcommand{\sd}[4]{\mathrm{SM}({#3}\lang_{#1},\lang_{#2}^{#4})}                        
\newcommand{\emm}[4]{\mathrm{EM}({#3}\lang_{#1},\lang_{#2}^{#4})}

\theoremstyle{definition}
\newtheorem{exmp}{Example}
\newtheorem{deff}[exmp]{Definition}
\Crefname{deff}{Definition}{Definitions}
\Crefname{exmp}{Example}{Examples}
\DeclareMathOperator*{\argmax}{arg\,max}
\DeclareMathOperator*{\argmin}{arg\,min}

\usepackage{flushend}
\title{Pragmatic Goal-Oriented Communications under Semantic-Effectiveness Channel Errors}

\author{\IEEEauthorblockN{Tomás Hüttebräucker, Mohamed Sana, Emilio Calvanese Strinati}

\IEEEauthorblockA{CEA-Leti, Université Grenoble Alpes, F-38000 Grenoble, France\\
Email: \{tomas.huttebraucker; mohamed.sana; emilio.calvanese-strinati\}@cea.fr}}

\newcommand{\titleheader}{This work has been accepted for publication in 2024 IEEE Consumer Communications and Networking Conference}

\makeatletter
\def\ps@IEEEtitlepagestyle{%
\def\@oddhead{\mbox{}\scriptsize \titleheader \rightmark \hfil }%
}
\makeatother
\begin{document}
\maketitle
\begin{abstract}
In forthcoming AI-assisted 6G networks, integrating semantic, pragmatic, and goal-oriented communication strategies becomes imperative. This integration will enable sensing, transmission, and processing of exclusively pertinent task data, ensuring conveyed information possesses understandable, pragmatic semantic significance, aligning with destination needs and goals. Without doubt, no communication is error free. Within this context, besides errors stemming from typical wireless communication dynamics, potential distortions between transmitter-intended and receiver-interpreted meanings can emerge due to limitations in semantic processing capabilities, as well as language and knowledge representation disparities between transmitters and receivers. The main contribution of this paper is two-fold. First, it proposes and details a novel mathematical modeling of errors stemming from language mismatches at both semantic and effectiveness levels. Second, it provides a novel algorithmic solution to counteract these types of errors which leverages optimal transport theory. Our numerical results show the potential of the proposed mechanism to compensate for language mismatches, thereby enhancing the attainability of reliable communication under noisy communication environments. 
\end{abstract}

\section{Introduction}

AI and beyond-5G communication systems need each other to grow but still follow separate paths. In AI algorithms, it is often assumed that communications are ideal, latency-power-error free, trustworthy, and always available (on demand). This is critically far from operational reality. On the other hand, 5G is not AI-native \cite{Calvanese6GVTM2019}[\cite{xiao2020selflearning}: Communication is designed to send raw data rather than AI understandable knowledge. While data is decisive to feed reasoning engines, being gathered-transferred-processed from scattered sources to update models, to accumulate knowledge and attain goals, communications are not engineered to be goal-oriented. The current approach is to encode data into symbols, ensuring symbol level accurate recovery even at ultra-high data rate exchange, regardless of their conveyed meaning \cite{Calvanese20216g} and agnostic to what cooperating reasoning agents need and can understand. Conversely, an AI-native approach requires to convey meaning or to accomplish a goal with a reasoning engine. Thus, what really matters is the impact that the received symbols have on the interpretation of the meaning intended by the transmitter and/or on the accomplishment of a common goal.
In this context a new impulse in research has started: \textit{semantic} and \textit{goal-oriented} communications have been proposed for future AI-native 6G networks \cite{Calvanese20216g} \cite{GeoffreyLi2021SemCom},\cite{Kountoris2021}, based on the seminal works of Shannon and Weaver \cite{shannon1950mathematical} and Bar-Hillel and R. Carnap \cite{bar-hillel1953semantic}, then on the more recent work of J. Bao \cite{bao2011towards}. This new area of research focuses on developing theories and algorithmic solutions for extracting, sharing and interpreting distilled information representing the significance of data, thus conveying the essential meaning to enable effective task execution. Such exchange of semantics that allows cooperation between intelligent agents requires a shared language which serves as a tool to endow real world observations with meaning that can be interpreted and thus enables decision making. The design of such language is a challenging task and it often relies on \ac{ml} algorithms. \ac{ml}-based language design has been extensively studied \cite{foerster2016learning}, \cite{lazaridou2018emergence}. Research indicates that incorporating communication as part of the learning procedure results in superior performance compared to human--designed communication protocols \cite{tung2021effective} \cite{bourtsoulatze2019deep}.

Most literature on the state of the art on semantic and goal oriented communication assumes that interacting agents share a common language. However, this might be an unrealistic assumption. Indeed, in future 6G communication multiple heterogeneous autonomous agents will communicate and collaborate, and they may utilize distinct languages. When agents use different languages, \textit{semantic noise} causes interpretation errors and performance degradation \cite{sana2022learning}. 
Similarly to \cite{sana2023semantic}, this work addresses the problem of communication errors due to language mismatch between a transmitter and a receiver. We model the communication channel in three levels, the syntactic (wireless) level, the semantic (meaning) level and the effectiveness (goal) level. A language deals with all three levels of the communication channel: it has to enable the completion of the task (\textit{effectiveness}), it has to allow for meaning transmission (\textit{semantic}) and it has to account for possible channel noise (\textit{syntactic}). In order to study heterogeneous language interactions, we treat a language mismatch as a semantic level problem and explore its consequences at the effectiveness level. 

The \textbf{contribution of the paper} is three-fold: (1) We extend the work from \cite{sana2023semantic} to language mismatch between agents in a \ac{mdp} context, (2) we propose a novel mathematical model for errors stemming from language mismatcht at both semantic and effectiveness levels and, (3) we propose an equalization algorithm that leverages optimal transport theory to counteract language mismatches and show its efficacy on a practical example.

\section{System Model}

\begin{figure}
    \centering
    \includegraphics[width=\linewidth]{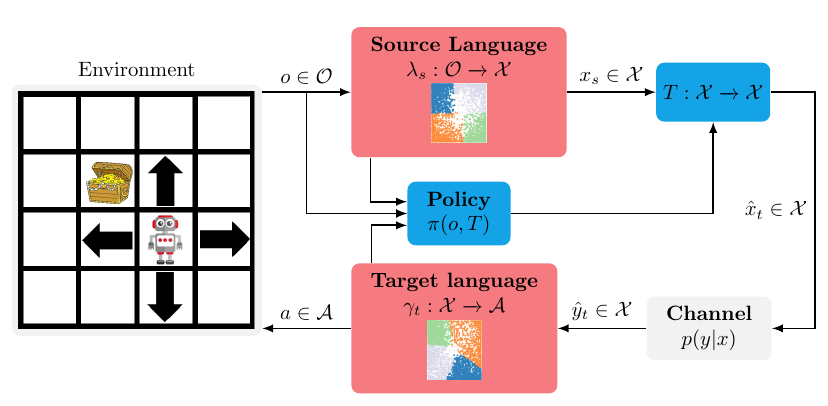}
    \caption{System model.}
    \label{fig:system_model}
\end{figure}

We will consider the system model shown in \cref{fig:system_model}. A transmitter observes a world, here the states of a stochastic environment, and encodes the underlying information for a distant receiver, which rules it to an action to complete a task. Communication takes place over an error-prone wireless channel, in several rounds, until the task is completed or it fails. To achieve the task, both encoder and decoder employ distinct languages (different logic), here artificially modeled with \ac{nn}, which we respectively refer to as \emph{language generator} and \emph{language interpreter}.

\noindent
\textbf{Language generator:}
A language generator $\enc$ maps an observation of the world into semantic representation: a latent representation capturing the intrinsic  structure of the data, which is useful for solving downstream tasks. We define such mapping with $\enc:\ospace\xrightarrow{}\sspace$, where $\ospace$ denotes the observation space and $\sspace$ the semantic space. Both $\ospace$ and $\sspace$ are topological spaces, we will assume $\ospace$ to be discrete and $\sspace$ to be continuous, however, our results are easily extended to different spaces. Thus, given an observation $\obs\in\ospace$, the language generator outputs the associated semantic representation $\sem\in\sspace$, later referred to as a semantic symbol. It is worth noting that $\lambda$ can, but is not required to be a one-to-one mapping. It can be a many-to-one or one-to-many mapping. In addition, $\lambda$ is not necessarily deterministic. It can be stochastic, in which case,  $\enc(\obs)=\sem\sim \mu_\enc(\sem'|\obs)$. Generally speaking, we can think of a deterministic mapping as a particular case of a stochastic one where $\mu_\enc(\sem|\obs)=\delta(\sem-\enc(\obs))$. Hence, given a distribution $\mu$ over the observation space $\ospace$, the language generator $\enc$ defines a distribution over the semantic space $\sspace$ as:
\begin{equation}\label{eq:sem_distribution}
    \mu_{\enc}(\sem)=\sum_{\obs\in\ospace} \mu_\lambda(\sem|\obs) \mu(\obs).
\end{equation}
A language generator defines a way to encode information from the observation space into the semantic space. This effectively creates a partition of the semantic space. A partition of the semantic space $\satom=\{\satom_1,\satom_2,\dots,\satom_N\}$ is a collection of measurable sets that cover $\sspace$, where the measure is $\mu_\enc$. The set $\satom_i\in\sspace$ is referred to as the $i$-th atom of the partition and $\mu_{\enc}(\satom_i)$ is its associated probability measure\footnote{Here, $\mu_{\enc}(\satom_i)$ can be interpreted as the probability that given a random observation $\obs$, $\sem=\lambda(\obs)$ belongs to $\satom_i$.}. Each atom $\satom_i$ is associated to a given semantic meaning. For example, in an image classification task, each atom is related to a different visual feature present of the images, meaning that if an image has a certain visual feature, its semantic representation will reside in the corresponding atom. In general, the semantic space partition depends only on the language generator and two different generators can partition the language differently for the same task. However, it has been observed that task specific optimal information encoding strategies usually share common characteristics \cite{moschella2022relative}.

\noindent
\textbf{Language interpreter:}
A language interpreter $\dec$ transforms a semantic representation into a task specific action in each communication round. More specifically, a language interpreter defines a strategy for interpreting the semantic partition defined by the language generator to solve the downstream task. We define such selection strategy with $\dec:\sspace\xrightarrow{}\aspace$, where $\aspace$ denotes the action space, which we will assume to be discrete. Thus, given a received semantic symbol $\sem\in\sspace$, the language interpreter selects an action $\act = \dec(x)\in\aspace$, possibly sampled from a probability distribution $\mu_\dec(\act'|\sem)$. In the image classification task example used for the language generator, the language interpreter leverages the semantic atom features of the generator's partition to infer a specific class. Hereafter, we denote a language (or a communication strategy) as a tuple $\lang=(\enc, \dec, \mu, \ospace, \sspace, \aspace)$.

\subsection{Emergence of language and communication effectiveness}
An effective communication between the encoder and decoder requires a joint learning procedure, during which a \emph{shared} language emerges to solve the downstream task. An efficient language generator should define a minimal representation, transparent and sufficient that enables the language interpreter to accomplish the task. A representation is sufficient if it encodes all the information relevant for the decision making and transparent if the encoding process is understandable by the interpreter, i.e. the encoding and decoding logic is the same. Minimality is related to the compression capabilities of the encoder and is directly linked to a particular task: a representation is minimal if it encodes only the task relevant information.

Ideally, the learned languages logic should only be affected by the task definition, the designed loss function and other architecture-specific constraints such as the characteristics of the semantic space $\sspace$. However, multiple other factors may affect the outcomes of the training such as the random weights initialization, training hyperparameters, or other sources of randomness including syntactic channel noise \cite{moschella2022relative}. As a result, even with the same \ac{nn} architecture and constrains, different training processes can yield different languages which nevertheless achieve similar performances on the task.
% Even though different languages define different communication strategies, this communication strategies may achieve similar performances. 
When the language generator and interpreter are the result of different training procedures, communication may suffer from language mismatch or incompatibility that stimulates effectiveness errors. This is the focus of our work, to define a novel framework to counteract and limit the critical errors in interpretation to achieve effective communication.

\subsection{Semantic channel and effectiveness channel}
In this section, we propose to model the semantic and effectiveness channel between an encoder-decoder pair, which do not share a common language. We denote with $\lang_\ls=\left (\enc_\ls, \dec_\ls, \mu, \ospace, \sspace, \aspace \right)$ the source language of the encoder and with $\lang_\lt=\left (\enc_\lt, \dec_\lt, \mu, \ospace, \sspace, \aspace \right)$ the target language of the decoder. We assume that the mismatch is only due to the language generator and interpreter, and that the encoder and decoder share the observation, semantic and action spaces as well as that the task that they haven been trained to solve. Let $\satom^\lin=\{\satom^{\lin}_1, \satom^{\lin}_2,\dots,\satom^{\lin}_N\}$ be the partition defined by the language generator $\enc_\lin$, $\lin\in\{\ls, \lt\}$. Also, let $\mathbbm{1}_{B}(\sem)$ denote the characteristic function on the subset $B\subset\sspace$ defined such that:
\begin{align}
    \mathbbm{1}_{B}(\sem) = {}\left \{
						\begin{array}{l l}
							1 & \text{~if~} \sem \in B,\\
							0 & \text{~if~} \sem \not\in B.
						\end{array}
					\right.
\end{align}
Given a semantic symbol $\sem\in\sspace$, we refer to $[\sem]_\lin$ as the atom to which $\sem$ belongs in the corresponding language, i.e. $\mathbbm{1}_{[\sem]_\lin}(\sem) = 1$. As the source and target language differ, an observation $\obs\in\ospace$ may be mapped to symbol $\sem_\ls=\enc_\ls(\obs)$ by the source language, and to symbol $\sem_\lt=\enc_\lt(\obs)$ by the target language. Semantic mismatch may arise if $\sem_\ls$ and $\sem_\lt$ do not belong to the same atom of the target language partition\footnote{Important note: $\sem_\ls$ and $\sem_\lt$ can differ, i.e. $\sem_\ls\neq \sem_\lt$ but they still belong to the same atom; there is no semantic error in this case as semantic is associated to the atoms of a partition.}. Hence, we introduce the following metric to quantify the semantic mismatch between source and target language: 
\begin{deff}[Semantic mismatch]\label{deff:semantic_distance}
 The semantic mismatch \normalfont SM \itshape between a source language $\lang_\ls$ and a target language $\lang_\lt$ is
    \begin{equation}
        \sd{s}{t}{}{}=1-\sum_{\obs\in\ospace} \mu(\obs) \int_{\sspace}\mathbbm{1}_{[\enc_\lt(\obs)]_t}(\sem)
                        \mu_\enc{_\ls}(\sem|\obs) d\sem 
    \end{equation}
\end{deff}
Intuitively, $\sd{s}{t}{}{}$ characterizes the probability of semantic misinterpretation between the two languages. Given an observation $\obs\in\ospace$, when the symbol $\sem_\ls\sim \mu_\enc{_\ls}(\sem|\obs)$ output by the source language does not belong to the corresponding semantic atom of the target language, the characteristic function is $\mathbbm{1}_{[\enc_\lt(\obs)]_t}(\sem_\ls) = 0$ which leads to semantic misinterpretation at the receiver. Note that we have $0\leq\sd{s}{t}{}{}\leq1$. When $\sd{s}{t}{}{}=1(0)$, $\mathbbm{1}_{[\enc_\lt(\obs)]_t}(\sem) = 0 (1)$ for all observations, indicating that no (all) semantic meaning is correctly conveyed. It is also worth noting that $\sd{s}{t}{}{}$ is not symmetric.

Errors at the semantic level do not necessarily translate to errors at effectiveness level. The effectiveness mismatch is a metric centred around the actions and how they affect the performance on the task. To capture how each action affects the task, we define, inspired by the the \ac{rl} literature, the action value function $\q:\aspace\times\ospace\xrightarrow{}\mathbb{R}$ which assigns to each observation-action pair a real number. The number $\q(\act,\obs)$ is called the ``value" of taking action $\act$ when the observed state is $\obs$ and it is related to the optimality of the action. If we interpret solving a task as equivalent to optimizing an objective function, the value $\q(\act,\obs)$ measures the effect of the decision on the objective function. The higher the value $\q(\act,\obs)$, the better that choosing action $\act$ in the context of $\obs$ is at optimizing the objective. So we should interpret $\q$ as a function that allows to find the best action for the current observation and to evaluate how sub-optimal other actions are. Using the function $\q$, we can then define the effectiveness mismatch as

\begin{deff}[Effectiveness mismatch]\label{deff:effectiveness_mismatch}
Given source language $\lang_s$ and target language $\lang_t$ and denoting  $\hat{\act}(\obs)=\dec_t(\enc_s(\obs))$ as the interpreted action for observation $\obs$ when the source encoder $\enc_s$ communicates with the target decoder $\dec_t$, the effectiveness mismatch can be measured by
    \begin{equation}
        \emm{s}{t}{}{}=1-\sum_{\obs\in\ospace}\mu(\obs)\frac{\q^+(\hat{\act}(\obs),\obs)}{\q^+(\act^*(\obs),\obs)},
    \end{equation}
where $\q^{+}(\act,\obs)=\q(\act,\obs)-\min_{\act'} \q(\act',\obs)$ is the positive-action value function for the task and $\act^*(\obs)=\argmax_{\act} \q(\act,\obs)$. Given that $0\leq\q^+(\act,\obs)\leq\q^+(\act_t^*(\obs),\obs)$ for all $\obs\in\ospace$ and $\act\in\aspace_t$ and that $\mu$ is a probability measure on $\ospace$, we have $0\leq\emm{s}{t}{}{}\leq 1$.
\end{deff}
The effectiveness mismatch is a normalized measure that captures how the performance on the task is degraded due to the language mismatch. When $\emm{s}{t}{}{}=0$, even if there is some language mismatch and subsequent semantic misinterpretations, the performance on the task is the optimal one. On the other hand, when $\emm{s}{t}{}{}=1$, the language mismatch is such that the performance is the worst possible one. As with the semantic mismatch, the effectiveness mismatch is not symmetric and depends on the choice of source and target languages. We note that, assuming that the target language leads to a behaviour that follows $\q$, $\emm{s}{t}{}{}\neq0$ implies that $\sd{s}{t}{}{}\neq 0$ (task performance degradation implies semantic misinterpretation), but $\sd{s}{t}{}{}\neq 0$ does not imply $\emm{s}{t}{}{}\neq0$ (misinterpretation does not imply task performance degradation). This is possible because multiple actions might achieve the optimal performance at a given time step and thus incorrect action interpretation might still lead to optimal performance. This is the principal distinction between semantic and effectiveness mismatch that will motivate the different equalization strategies developed in this work.

\section{Semantic Channel Equalization}

In this section we will introduce the concept of semantic channel equalization through \textit{measurable transformations}. A measurable transformation $\mt$ is a function that transforms values measured by a probability distribution into values measured by another probability distribution. This way, by using $\mu_{\enc_\ls}$ and $\mu_{\enc_\lt}$ as measures over $\sspace$, a transformation $\mt:\sspace\xrightarrow{}\sspace$ is a measurable transformation.
% is a measurable transformation since $\mu_{\enc_s}$ and $\mu_{\enc_t}$ measure $\sspace$.
We call this type of transformation a \textit{semantic equalizer}. Semantic equalization between a source language $\lang_s$ and a target language $\lang_t$ is the process of transforming, for each observation $\obs$, a source semantic symbol $\sem_s=\enc_s(\obs)$ into $\hat{\sem}_t=\mt\circ\enc_s(\obs)$. We denote the set of measurable transformations of $\sspace$ onto itself as $\mathcal{\mt}$. We can then define a language equalization,

\begin{deff}[Semantic language equalization] Let $\lang=(\enc, \dec, \mu, \ospace, \sspace, \aspace)$ be a language and $\mt\in\mathcal{\mt}$ be a semantic equalizer, the equalized version of $\lang$ is 
\begin{equation}
    \mt\lang=(\mt\enc, \dec, \mu, \ospace, \sspace, \aspace)
\end{equation}
where $\mt\enc=\mt\circ\enc$ denotes the transformed language generator.
\end{deff}
In this work we deal with equalization from the side of the encoder, hence the modified language generator $\mt\enc$ in the definition of equalized language, but notice equalization can be performed at the decoder as well.

The optimal semantic equalizer is the one that minimizes either the semantic or the effectiveness mismatch between the transformed source language and the target language. It can be found by solving the optimization problem
\begin{equation}\label{eq:main_opt}
    \mt^*=\argmin_{\mt\in\mathcal{\mt}}\left[ f(\mt\lang_s,\lang_t) \right].
\end{equation}
Where $f(\mt\lang_s,\lang_t)=\sd{s}{t}{\mt}{}$ if the objective is to minimize the semantic mismatch and $f(\mt\lang_s,\lang_t)=\emm{s}{t}{\mt}{}$ if the objective is to minimize the effectiveness mismatch. We aim at finding the optimal semantic equalizer $\mt^*$. Solving the problem in \cref{eq:main_opt} is challenging and a transformation that can correctly map the whole source semantic space into the target semantic space can be highly complex. Therefore, we propose to design a \textit{codebook} of multiple low complexity transformations that, together with a selection policy, can effectively equalize the semantic channel.

\subsection{Transformation Codebook}
We focus now on languages with finite discrete action spaces, which makes the design of a finite and discrete transformation codebook feasible. In this case, we can assign to each action a semantic atom, and thus the semantic partition will also be discrete and finite. The transformation codebook we plan to design consists of multiple low-complexity linear transformation that map semantic atoms from the source language into semantic atoms of the target language. Let $\aspace_\lin =\{\act^\lin_i|i\in J_\lin\}$ be the action set of language $\lang_\lin$, where  $J_\lin = \{1,2,\dots,K_\lin\}$ denotes the action set index and let $\satom^{\lin}_i$ be the atom of the semantic space corresponding to action $\act_i^l$ of language $\lang_l$. The codebook of transformations can be designed by solving the following problem

\begin{equation}\label{eq:codebook_opt}
\begin{split}
    \cb_{\mct} =  \biggl\{ \mt_{i,j}=\argmax_{\mt\in\mct} \left [ \infotransfer{i}{j}(\mt) \right ]  \forall i \in J_s, j\in J_t  \biggr\}
\end{split}
\end{equation}
where
\begin{equation}\label{eq:information_transfer}
\begin{split}
    \infotransfer{i}{j}(\mt) =   \frac{\mu_{\mt\enc_s}\left(\mt\left (\satomsource{i} \right )\cap \satomtarget{i}\right)}{\mu_{\mt\enc_s}\left(\satomsource{i}\right)} 
\end{split}
\end{equation}
is the information transfer between source semantic atom $\satomsource{i}$ and target semantic atom $\satomtarget{j}$ under the semantic equalizer $\mt$ \cite{sinha2019Information}. Here, we define $\mt(A)=\{\mt(\sem)|\sem\in A\}$. Since $\mu_{\mt\enc_s}$ is the equalized source distribution on the semantic space, $\infotransfer{i}{j}(\mt)$ measures the probability that an element from $\satomsource{i}$ falls in $\satomtarget{j}$ after being equalized by $\mt$. We also have that $0\leq\infotransfer{i}{j}(\mt)\leq1$.

\subsection{Semantic and effectiveness channel equalization using optimal transport}
This work leverages \ac{ot} to solve the problem in \cref{eq:codebook_opt}. \ac{ot} is a method to map a set of samples from a source distribution into samples from a target distribution while minimizing the transportation cost (e.g. euclidean distance). This effectively means transforming samples from the source distribution in a way that they appear to be sampled from the target distribution. In particular, in \cite[Sec. 3.2]{perrot2016mapping} a method to obtain a linear approximation of the optimal transportation map is presented. By using this method, we show that it is possible to compensate non linear language mismatch by operating a codebook of \textit{low-complexity linear-transformations}. For more details on the optimal transport algorithm used to obtain the codebook we refer the reader to \cite{sana2023semantic}.

\subsection{Policy for transformation selection}
To complete the semantic equalization method, once $\cb_{\mct}$ is obtained, a policy $\pi$ that selects a transformation from  $\cb_{\mct}$ and applies it on a given semantic symbol needs to be designed. The selection policy can be designed to either minimize the semantic mismatch (\cref{deff:semantic_distance}) or the effectiveness mismatch (\cref{deff:effectiveness_mismatch}). For each observation $\obs\in\ospace$, the encoder has to select $\mt\in\cb_{\mct}$ to pre-equalize his message following a policy $\pi$ that will minimize the desired metric. We will focus on finding a probabilistic policy $\pi$ where $\pi(\obs,\mt)$ specifies the probability of choosing transformation $\mt$ to equalize $\enc_s(\obs)$. We define
\begin{equation}\label{eq:policy_opt}
    \pi(\obs,\mt) = \argmin_{\hat{\pi}\in\Pi} \mathbb{E}_{\obs\sim\mu(\obs)} \left[R(\hat{\pi},\obs)\right]
\end{equation}
where $R\in\{R_S,R_E\}$ is a risk measure that can either be the \textit{semantic mismatch} risk or the \textit{effectiveness mismatch} risk, both this risk functions will be detailed next. .

\subsubsection{Semantic mismatch risk}
To minimize the semantic mismatch, a selection policy should choose the transformation $\mt$ that maps each semantic $\sem_\ls\in\satomsource{i}$ into a corresponding target symbol $\hat{\sem}_\lt\in\{ \satomtarget{j}|j\in\kappa(i)\}$. Here $\kappa:J_s\xrightarrow{}J_t$ captures the correspondence between source and target semantic atoms. $\kappa$ can be a one-to-one mapping, many-to-one or one-to-many depending on the source and target language generators, the action space and the downstream tasks to solve by both languages. In our scenario, both the source and the target language were designed to solve the same task with the same action space, so we can find a one-to-one correspondence between atoms, i.e. atoms are associated if they lead to the same action. We then define the semantic mismatch risk as
\begin{equation}\label{eq:sem_risk}
\begin{split}
    R_S(\pi,\obs) =  1-\mathbb{E}_{\mt\sim\pi(\obs,\mt)}&\Bigg[ \sum_{i\in J_s} \mu_{\enc_s}\left (\satomsource{i}|\obs \right) \\ 
    & \sum_{j\in\kappa(i)} \infotransfer{i}{j}(\mt) \Bigg ]
\end{split}
\end{equation}
where $\mu_{\enc_s}\left (\satomsource{i}|\obs \right)$ is the probability that the semantic symbol $\sem_\ls=\enc_s(\obs)$ belongs to the semantic atom $\satomsource{i}$. 

\subsubsection{Effectiveness mismatch risk}
Not all semantic misinterpretations lead to the same degradation in performance, even more, it is possible that, after experiencing a semantic error, the effectiveness aspect of the system isn't affected at all. This is the main motivation behind reducing the effectiveness mismatch rather than the semantic mismatch. Denoting as $\q_t(\act,\obs)$ the target language estimation of the true $\q(\act,\obs)$ value function, the effective mismatch risk is 
\begin{equation}\label{eq:eff_risk}
\begin{split}
    R_E(\pi,\obs) =  1-&\mathbb{E}_{\mt\sim\pi(\obs,\mt)}\Bigg[ \sum_{i\in J_s} \mu_{\enc_s}\left (\satomsource{i}|\obs \right) \\ 
    & \sum_{j\in J_t} \infotransfer{i}{j}(\mt)\cdot \q_t(\act_j,\obs) \Bigg ]
\end{split}
\end{equation}
where $\act_j=\dec_t(\satomtarget{j})$ is the action the target decoder outputs for any semantic message inside partition $\satomtarget{j}$. Notice that, since it relies on the estimation $\q_t(\act,\obs)$, this equalization strategy is tied to the performance of the target language.

\section{Numerical results}
We test the proposed system on a language designed by \ac{rl} methods to solve the environment shown in \cref{fig:system_model}. In this scenario, the observation space $\ospace$ is the state of a grid world with an agent and a treasure. The encoder $\enc$ maps each observation into $\sspace=\mathbb{R}^2$ in a deterministic manner and transmits it through a AWGN channel. The output power of the encoder is normalized o that that each of the components of the semantic symbol $\sem=(\sem_1,\sem_2)$ is limited to 1, i.e. $\sem_i\leq1, i\in\{1,2\}$. This way, since it is the maximum and not the mean signal power which is fixed, the characteristics of the channel are defined by the \ac{psnr} rather than by the  \ac{snr}. However, to simplify notation, from now on we will denote the \ac{psnr} as \ac{snr}. The decoder $\dec$ receives the noisy version of the transmitted symbol and chooses an action from
% in the action space
$\aspace=\{\text{right, down, left, up}\}$ which is then executed by the agent. For the action taking, we will explore both a deterministic decoder and a stochastic one. The objective is to get the agent to the treasure in the least number of time steps possible. The episodes finishes when the agent gets to the treasure or when the maximum number of steps (150) is achieved.

\textit{We assume that the encoder and decoder are given and that they were jointly trained in a de-centralized manner using \ac{rl} techniques with the same reward signal with a \ac{snr} of 10 dB}. The decoder is trained to learn $\q(\act,\sem)$ using Deep Q-Learning (DQN) \cite{mnih2015human} which is suitable for discrete action spaces while the encoder is trained using Deep-Deterministic policy gradient (DDPG) \cite{lillicrap2015continuous} which is suitable for a continuous action space. For more details on the training we refer the reader to \cite[Section IV]{tung2021effective}. 

During evaluation time, the decoder decision making is defined as
\begin{equation}\label{eq:decoder_softmax}
    \dec(\sem) = \act  \sim \textrm{SoftMax}\left (  \beta \q(\act,\sem) \right )
\end{equation}
where $\beta$ is the inverse of the well known temperature value for a soft max distribution where $\beta=0$ corresponds to completely random decisions and the deterministic case can be recovered by setting $\beta\xrightarrow{}\infty$. For the results we will explore both a determinist decode and a stochastic one with $\beta=5$.

\subsection{Emergent language diversity}

\begin{figure}
    \centering
    \includegraphics[width=\linewidth]{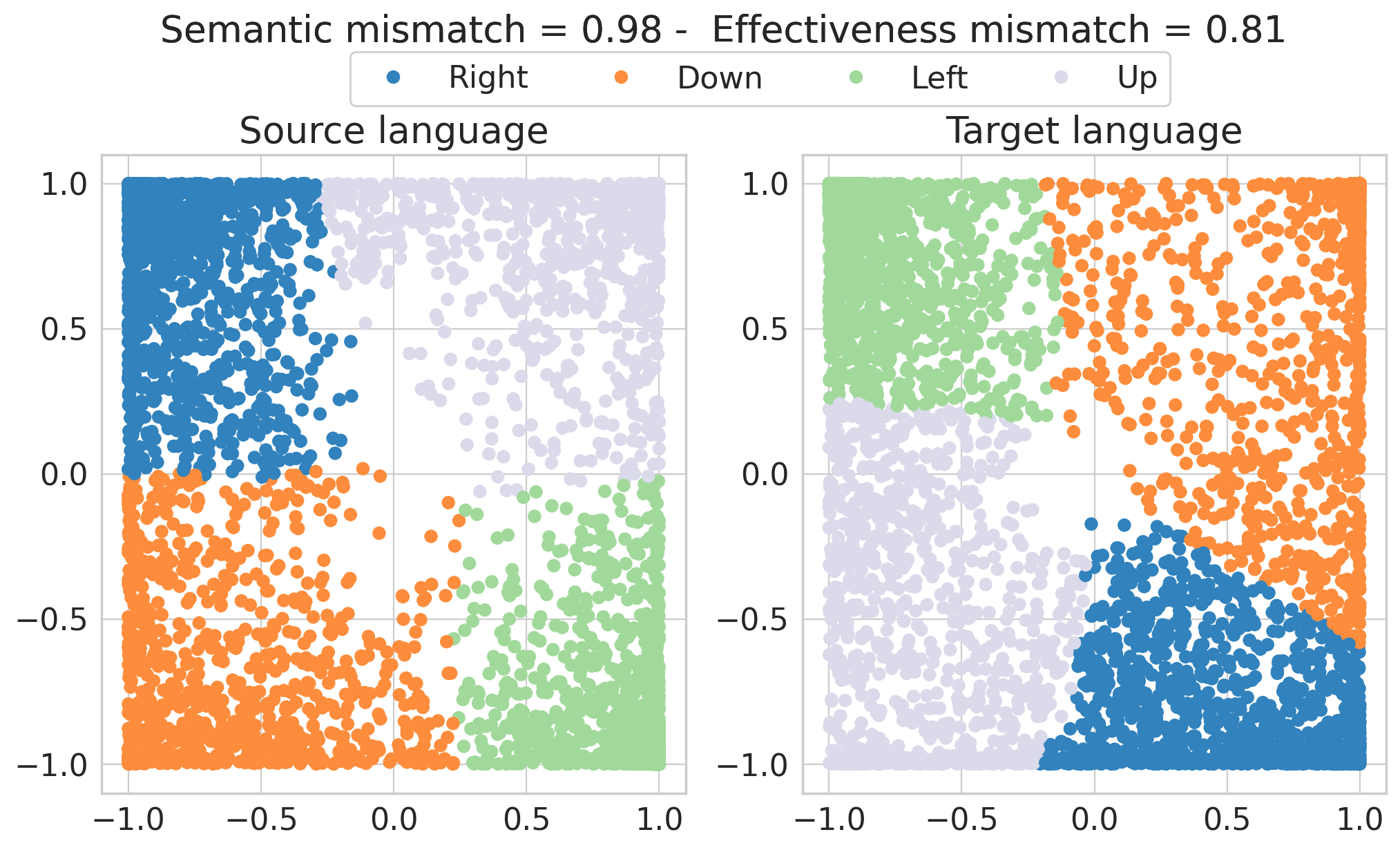}   
    \caption{Mapping of the learned languages. Both encoder-decoder pairs were trained to solve the same task in identical conditions only differing in stochastic training aspects.}
    \label{fig:language_mapping}
\end{figure}

In \cref{fig:language_mapping} the semantic language partition is shown for both the target and the source language. We see that each language follows a logic that divides the semantic space in different ways with a successful task competition. The training algorithm, \ac{nn} architectures and training task are identical for both languages, this shows that the training results are sensible to random weight initialization and randomness in training. In particular we see that opposite actions (up-down and right-left) are mapped the furthest away allowed by the available semantic space (which is limited by the power normalization) for both languages. This characteristic suggests that, even if multiple optimal semantic space partitions are possible, there is a certain structure that allows this languages to successfully complete a task. This latent space invariability has been observed for non distributed \ac{ai} in multiple task \cite{moschella2022relative}.

\subsection{Language equalization}
\begin{figure}
    \centering
    \includegraphics[width=\linewidth]{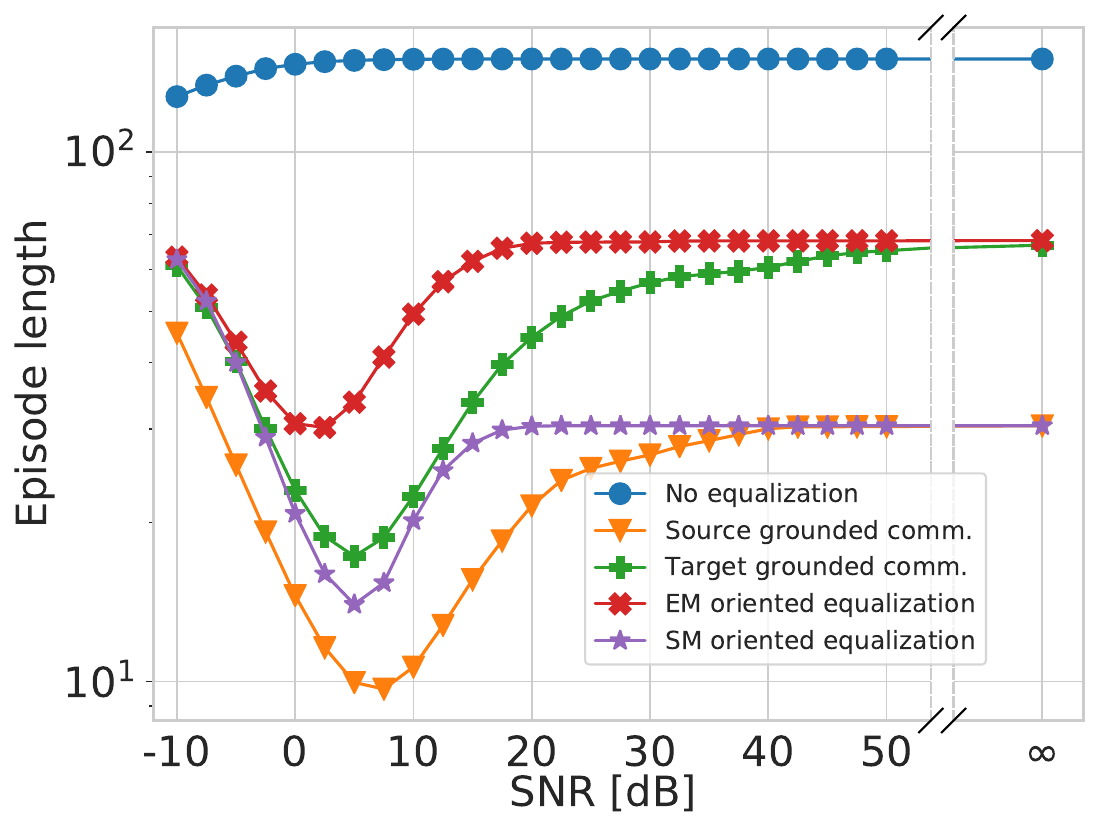}
    \caption{Average episode length (lower is better) for the different communication strategies with varying \ac{snr} for a deterministic decoder ($\beta\xrightarrow{}\infty$ in \cref{eq:decoder_softmax}).}
    \label{fig:performance_deterministic}
\end{figure}

\begin{figure}
    \centering
    \includegraphics[width=\linewidth]{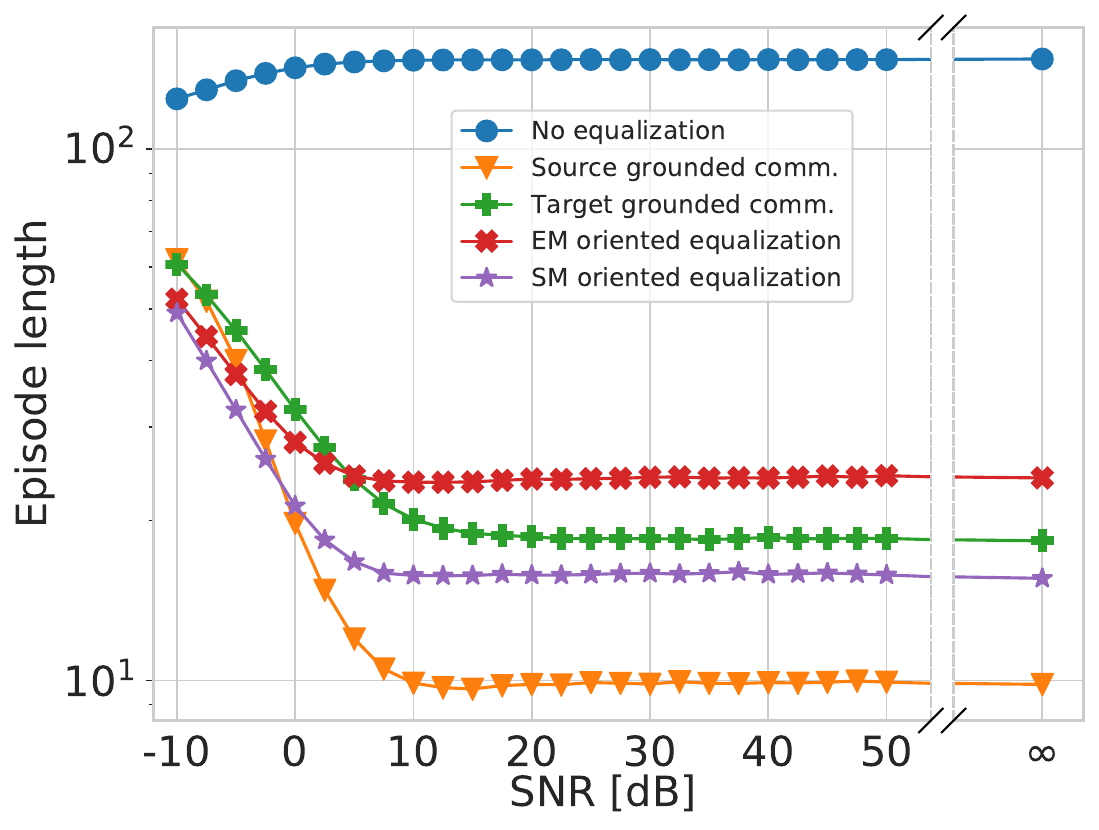}
    \caption{Average episode length (lower is better) for the different communication strategies with varying \ac{snr} for a stochastic decoder with $\beta=5$ in  \cref{eq:decoder_softmax}.}
    \label{fig:performance_stochastic}
\end{figure}

The metric to evaluate the performance of the system is the episode length. The objective of the communication is for the agent to find the treasure as fast as possible, thus, lower episode length means better performance. We compare the performance of the proposed methods with other approaches both for a deterministic decoder (see \cref{fig:performance_deterministic}) and a stochastic one (see \cref{fig:performance_stochastic}). The evaluated communication strategies are:

\indent \textbf{\textit{No equalization}:} The source encoder messages are not equalized i.e. the messages are transmitted as the encoder outputs them without further processing. We see that both for a deterministic (\cref{fig:performance_deterministic}) and a stochastic (\cref{fig:performance_stochastic}) decoder the performance of the system is worst. It seems counter intuitive to see that for low \ac{snr} values, the performance actually improves. This is due to the fact that the errors caused by the language mismatch can sometimes be corrected by the channel noise. 

\indent \textbf{\textit{Source/Target grounded communications}}: The encoder and decoder both adopt the source/target language, the two ideal cases with no mismatch. For the deterministic case, \cref{fig:performance_deterministic}, the languages attain the optimal performance in the \ac{snr} range from 0 to 10 dB rather than $\ac{snr}=\infty$. This is related to the limitation of the learned languages. Indeed, in the absence of channel noise, the encoder-decoder pairs do not always perform optimally. In fact, we have
% it has been 
observed that the \textit{language systematic error} causes the system to enter in loops that lead to task failing. This loops occur when the actions of the decoders bring them to a state that has already been visited before. In this situation, it is easy to see that, without any channel noise, the agent is doomed to follow the same steps as before and the system will end up transitioning between previously encountered states. Eventually, this will lead to the agents attaining the maximum time-steps and fail the task. It is the presence of noise that can break this type of infinite loops caused by the systematic error. This is further validated by the results for stochastic decoder shown in \cref{fig:performance_stochastic}. When we introduce some randomness in the decision making, the performance is not degraded for increasing \ac{snr} since the infinite loops in the episode can be avoided. In this case, channel noise is actually detrimental and the episode length is minimized for infinite \ac{snr}. This suggests that, in the deterministic decision-making scenario, reducing the syntactic channel noise may result in the increase of the semantic noise, which affect the performance in achieving the goal. In contrast, when considering the stochastic decoder (\cref{fig:performance_stochastic}), reducing the syntactic noise improves the goal achievement performance. These results show the strong interplay between syntactic noise and semantic noise and how they affect each other.

\indent \textbf{SM \itshape and \upshape EM \itshape oriented equalization:} The proposed \ac{ot} based equalization method. For SM oriented equalization, the policy that minimized the semantic mismatch risk in \cref{eq:sem_risk} is used. For EM oriented equalization, we employ a policy that minimizes the effectiveness mismatch risk in \cref{eq:eff_risk}. Notice that, since it follows $\mu_{\enc_\ls}$, the SM oriented equalization is tied to the source language performance. On the other hand, the EM oriented equalization is tied to the target language performance by the target language Q function $\q_t$.

In both the cases of a deterministic (\cref{fig:performance_deterministic}) and a stochastic (\cref{fig:performance_stochastic}) decoder, our methods are effective on minimizing the language mismatch effects on the performance since the performance is improves for all values of \ac{snr} when compared to the \textit{no equalization} approach. However, the equalization methods perform worse than the base languages. This can be explained by the resilience of the proposed methods to channel noise, indeed, as the developed methods reduce the effect of the wireless channel noise, this enforces the systematic errors of the language and causes an increase in the average episode length. This hypothesis is further reinforced by the fact that, under error-free syntactic channel ($\ac{snr}=\infty$) and deterministic decoding (\cref{fig:performance_deterministic}), the performance of the SM and EM oriented equalization decreases and aligns perfectly with the source and target grounded communications respectively.

\subsection{Ablation study on $\beta$}
\begin{figure}
    \centering
    \includegraphics[width=\linewidth]{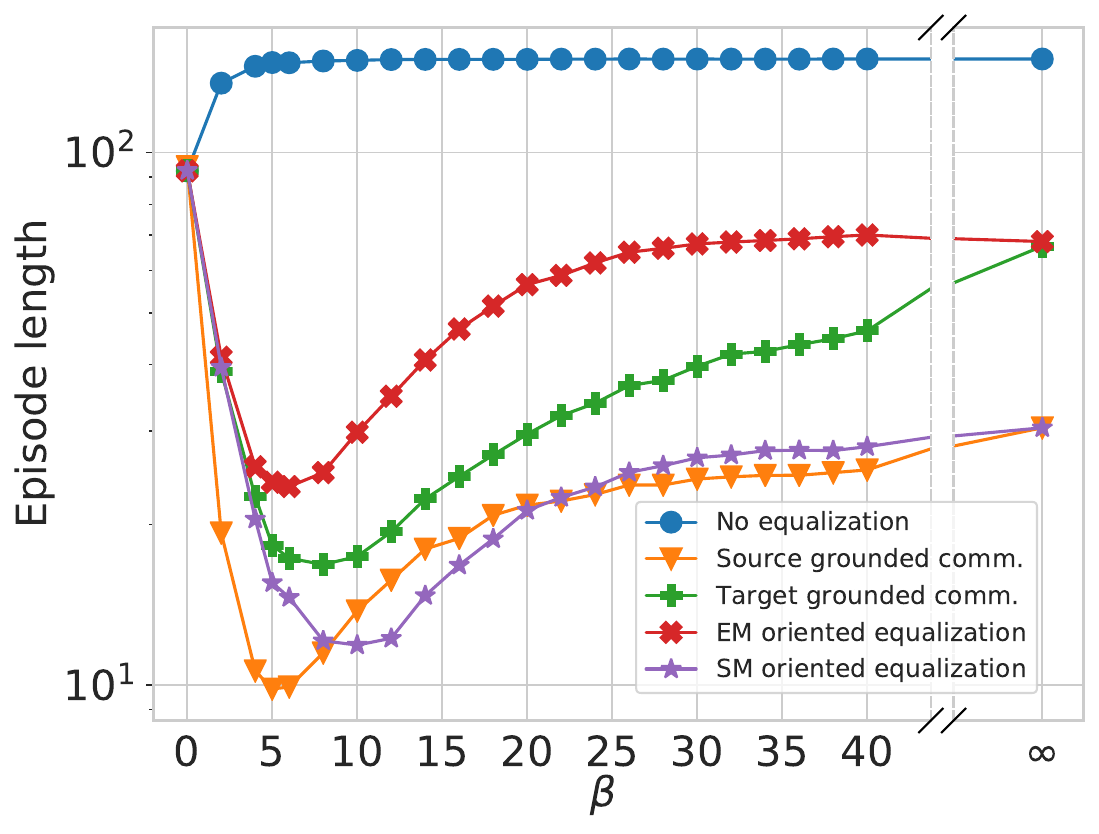}
    \caption{Average episode length (lower is better) for the different communication strategies with varying $\beta$ and $\ac{snr}=\infty$.}
    \label{fig:performance_beta}
\end{figure}

To better characterize the effects of an stochastic decoder, we performed a study of the performance of the different methods while varying $\beta$ and fixing $\ac{snr} =\infty$. In \cref{fig:performance_beta} the results are shown. We see that increasing $\beta$ has similar effects as increasing the \ac{snr} for the deterministic encoder (\cref{fig:performance_deterministic}). Indeed, as $\beta$ increases, the decoder becomes more and more deterministic to choose actions based on the maximum the $\q$-values principle (see \cref{eq:decoder_softmax}). Conversely, when $\beta$ decreases, the decoder becomes more stochastic. In particular, for $\beta=0$, all the decisions are random and thus independent from the communication, we see in this case that all the communication methods attain the same performance.

\section{Conclusions}

In this paper, we analyze the consequences of semantic extraction and interpretation mismatches on goal-oriented task solving in scenarios involving heterogeneous AI agents utilizing distinct languages. At first, we establish mathematical models that address errors originating from language mismatches, encompassing both semantic and effectiveness levels, employing measurable transformations within semantic representation spaces. We show that in case of language mismatch, emergent languages might encounter systematic errors in task solving. In order to assess and address this critical issue, we present innovative metrics aimed at quantifying the influence of language mismatch across both semantic and effectiveness dimensions.
  
Our proposed solution incorporates these metrics to devise an innovative equalization algorithm firmly grounded in optimal transport theory. This algorithm serves as a crucial component in mitigating systematic errors stemming from language mismatches. Notably, our proposed equalization methods effectively diminish language-related discrepancies, thereby enhancing the overall robustness of task execution. Furthermore, we investigate the behaviour of the system and validate the performance of our equalization methods when stochastic decision-making is used. This helps to understand the role that randomness and errors play in distributed \ac{mdp} problems.

\section*{Acknowledgements}
The present work was supported by the EU Horizon 2020 Marie Skłodowska-Curie ITN Greenedge (GA No. 953775), by the 6G-GOALS Project under the HORIZON program (GA No. 101139232), and by the French project funded by the program "PEPR Networks of the Future" of France 2030 (ref. 22-PEFT-0010).

\balance

\bibliographystyle{ieeetr}
\bibliography{bib}

\end{document}